\documentclass[twocolumn,showpacs]{revtex4}

\usepackage{graphicx}
\usepackage{dcolumn}
\usepackage{bm}
\usepackage{hyperref}
\usepackage{amsmath}
\usepackage{color}
\usepackage[utf8]{inputenc}

\begin{document}

\title{Binary black hole simulation with an adaptive finite element method \uppercase\expandafter{\romannumeral3}: Evolving a single black hole}

\author{Li-Wei Ji$^{1,2}$, Rong-Gen Cai$^{2}$, Zhoujian Cao$^{1}$}
\affiliation{$^1$Department of Astronomy, Beijing Normal University, Beijing, 100875, China}
\affiliation{$^2$CAS Key Laboratory of Theoretical Physics, Institute of Theoretical Physics,
	Chinese Academy of Sciences, Beijing 100190  
	and School of Physical Sciences,
	University of Chinese Academy of Sciences, Beijing 100049, China}

\pacs{
04.25.Dm,	
04.70.Bw,	
95.30.Sf	
%
}


%

\newcommand\p{{\partial}}

%

\begin{abstract}
We extend a new finite element code, Einstein PHG (iPHG), to solve the evolution part of Einstein equations in first-order GH formalism. This paper is the third one of a systematic investigation of applying adaptive finite element method to the Einstein equations, especially binary compact objects simulations. The primary motivation of this work is to evolve black holes for the first time utilizing a continuous Galerkin finite element method on unstructured (tetrahedral) mesh.   We test our code by evolving a nonlinear scalar wave equation. It works well and runs stably with both reflect and radiative boundary conditions. Then we use iPHG to simulate the full three-dimensional spacetime of a single black hole. We find that the filter used to dealt with aliasing error is a crucial ingredient for numerical stability. For simplicity, we impose the  ``freezing"  ingoing characteristic fields condition in weak form at the outer boundary. Our simulations show both the convergence and stability.
\end{abstract}

\maketitle

\section{\label{sec:level1_1}Introduction}
There exist three different kinds of numerical methods in the world: the finite difference method, the spectral method, and the finite element method. In the numerical relativity community, the first two methods are much more popular than the last one. In \cite{Aylott:2009ya,Cao:2008wn,Yamamoto:2008js,Clough:2015sqa}, finite difference method has been used to simulate coalescing of compact binaries. Spectral method has also been successfully used in \cite{Hilditch:2015aba,Hilditch:2017dnw,Lindblom:2005qh,Scheel:2008rj,Deppe:2018uye} to study gravitational collapse and binary black hole dynamics. However, a full three-dimensional binary black hole simulation using finite element method, which includes the whole inspiral-merger-ringdown phase, is still missing (but see \cite{Sopuerta:2005gz,Sopuerta:2005rd,Zumbusch:2009fe,Field:2010mn,Brown:2012me,Miller:2016vik,Dumbser:2017okk,Hebert:2018xbk}).

Even though we are now capable of studying all kinds of physical phenomena related to the strong gravitational field and highly dynamical spacetime, through numerical relativity \cite{Cardoso:2014uka,Baiotti:2016qnr}, there are still some challenges. For example, the gravitational waves calculated using existing finite difference or spectral codes are already accurate enough to make a detection in the network of laser interferometric detectors, such as LIGO and VIRGO. But the highest mass ratio of a binary black hole system, which can be successfully simulated now, is around 1:20. And it is not practical to use existing codes to simulate the sources with mass ratio far beyond this range. However, these high mass ratio binaries are expected to be observed by space-based interferometers such as LISA, Taiji, and TianQin. 

The reason why those finite difference or spectral codes are not suitable to simulate intermediate mass ratio inspirals (IMRIs), whose mass ratio is far beyond 1:20, is their pool parallel scalability. Large scale difference due to large mass difference requires large size parallel computing. So strong parallel scalability is essential for this kind of binary system simulation. For finite difference codes, especially for those who use moving-box mesh refinement techniques, it is their hierarchical structure that limits the parallel scalability. Furthermore, the size of the buffer zone due to the structure of finite difference's stencil also sets an up limit for its ability of parallel scaling. For (multi-domain) spectral codes, in principle, they can have strong parallel scalability. But it requires a lot of fine-tuning and complicated grid structures, which make  IMRIs simulations using spectral method quite challenging.

For finite element method, its discretization admits a local property similar to finite difference case, so its robustness is expected to be as good as finite difference method. While in each element, high order polynomial function basis and/or spectral function basis can be used to achieve high accuracy, just as the spectral method. Furthermore, in contrast to finite difference method where data has to be transferred between different mesh levels, all elements in finite element method are treated uniformly, which would make finite element method admit higher strong parallel scalability.

The application of finite element method to general relativity is just in the beginning and still needs a lot of exploration. Discontinuous Galerkin finite element method was implemented in \cite{Kidder:2016hev} to deal with general relativistic hydrodynamics and in \cite{Hebert:2018xbk} to evolve a Kerr black hole and a neutron star. Scott et al. \cite{Field:2010mn,Brown:2012me} used local discontinuous Galerkin finite element method to solve spherically reduced Baumgarte-Shapiro-Shibata-Nakamura (BSSN) system with first-order and second-order operators. Dumbser et al. \cite{Dumbser:2017okk} presented some preliminary results on the evolution of binary black-hole systems which is performed in a high-order path-conservative arbitrary-high-order-method-using-derivatives (ADER-DG) scheme. In \cite{Cao1}, we also used local discontinuous Galerkin finite element method to solve spherically reduced first order general harmonic (GH) system.   

Previously we have developed a new finite element code, iPHG \cite{Cao:2015via}, to solve the constraint part of the Einstein equations in full three dimensions. In the current work, we extend our code such that it can solve the evolution part of the Einstein equations. Our code is based on one recently developed adaptive finite element library--Parallel Hierarchical Grid (PHG) \cite{zhang2009parallel,zhang2009set}. PHG is a toolbox for writing scalable parallel adaptive finite element programs and provide functions which perform common and difficult tasks in parallel adaptive finite element programs, such as management of unstructured parallel (distributed) meshes, parallel adaptive mesh refinement and coarsening, dynamic load balancing via mesh repartitioning and redistribution and so on.

Throughout this work, the geometry units with $G = c = 1$ are used. The rest of the paper is arranged as follows. In the next section, we will briefly review the GH formalism of Einstein equations and introduce the numerical algorithm implementing them with the finite element method. Then in Sec. \ref{sec:level1_3}, we present the numerical results, including some simple tests and evolutions of single Schwarzschild black hole spacetime. We conclude in Sec. \ref{sec:level1_4}.

\section{\label{sec:level1_2}Numerical algorithm}
In the first-order reduction of GH formulation \cite{Lindblom:2005qh}, the state vector is denoted as $u^\alpha=\left\{g_{ab},\Pi_{ab},\Phi_{iab}\right\}$, where $g_{ab}$ is the spacetime metric, $\Pi_{ab}=-t^c\partial_cg_{ab}$ and $\Phi_{iab}=\partial_i g_{ab}$ are two auxiliary variables. Then the dynamical equations which are reduced from the Einstein equations can be written as
\begin{align}
\partial_tu^{\alpha}+A^{k\alpha}{}_\beta\partial_k{u^\beta}=S^\alpha, \label{eq: Einsteins}
\end{align}
with 
\begin{align}
A^{k\alpha}_\beta&=
\begin{pmatrix}
-(1+\gamma_1)\beta^k & 0 & 0 \\
-\gamma_1\gamma_2\beta^k & -\beta^k & \alpha \gamma^{ik} \\
-\gamma_2\alpha\delta^k_i       & \alpha\delta^k_i & -\beta^k
\end{pmatrix},
\end{align}
\begin{align}
S^\alpha&=
\begin{pmatrix}
-\alpha\Pi_{ab}-\gamma_1\beta^i\Phi_{iab} \\
S^{(\Pi)}_{ab}  \\
\alpha\left[\frac{1}{2}t^ct^d\Phi_{icd}\Pi_{ab}+\gamma^{jk}t^c\Phi_{ijc}\Phi_{kab}-\gamma_2\Phi_{iab}\right]
\end{pmatrix},
\end{align}
and 
\begin{align}
\begin{split}
S^{(\Pi)}_{ab}=&2\alpha g^{cd}(\gamma^{ij}\Phi_{ica}\Phi_{jdb}-\Pi_{ca}\Pi_{db}-g^{ef}\Gamma_{ace}\Gamma_{bdf}) \\
&-2\alpha\nabla_{(a}H_{b)}-\frac{1}{2}\alpha t^ct^d\Pi_{cd}\Pi_{ab}-\alpha t^c\gamma^{ij}\Pi_{ci}\Phi_{jab} \\
&+\alpha\gamma_0\left[2\delta^c_{(a}t_{b)}-g_{ab}t^c\right]C_c-\gamma_1\gamma_2\beta^i\Phi_{iab},
\end{split}
\end{align}
where $H_a$ is the source function for generalized harmonic formalism, $t^a$ is the unit vector normal to the spatial slices of constant coordinate time $t$, $\Gamma_a=g^{bc}\Gamma_{abc}$ are the contracted Christoffel symbol and $C_c=\Gamma_c+H_c$ is the constraint that ensures the coordinates satisfy the GH coordinate condition. The lapse $\alpha$, shift $\beta^i$ and spatial metric $\gamma_{ij}$ are defined by
\begin{align}
ds^2=&g_{ab}dx^adx^b \nonumber\\
=&-\alpha^2dt^2+\gamma_{ij}(dx^i+\beta^idt)(dx^j+\beta^jdt).
\end{align}
The terms multiplied by $\gamma_{0,1,2}$ are the additional constraint terms beyond original Einstein equations. In the simulations, we set $\gamma_0=\gamma_2=1, \gamma_1=-1$, as in \cite{Lindblom:2005qh}. Throughout this paper, we use the Latin $a,b,c,...$ for four-dimensional indices, while $i,j,k,...$ for spatial indices.

In the following, we will use the continuous Galerkin (CG) finite element method to solve Eq.~\eqref{eq: Einsteins}. We multiply it by a test function $v$ and integrate over the whole computational domain $\Omega$. Using integration by parts, we get the following weak form of the original equation, (here we omit the state index for simplicity,)
\begin{align}
\begin{split}
\int_\Omega \partial_tu vd^3x =&-\int_\Omega \partial_kA^{k}u vd^3x-\int_\Omega A^{k}u\partial_kvd^3x \\
&+\oint_{\partial\Omega} A^{k} uv n_k\frac{1}{\sqrt{\gamma}}d^2\Sigma +\int_\Omega S v d^3x \label{eq: weak1}
\end{split}
\end{align}
where $n_k$ is the outward directed unit normal to the boundary of $\Omega$, $d^2\Sigma$ is the invariant surface element \cite{Teukolsky:2015ega}.

Denoting the basis functions of the finite element $\phi_i$, we can expand the state vector $u$ as $u=\sum_i u_i\phi_i$. The set of test functions $v$ is also chosen to be the same with the set of basis functions (Galerkin method). Then the above weak form equations can be written as
\begin{align}
\begin{split}
M_{ij} \partial_tu_i=&-\int_\Omega \partial_kA^{k} u \phi_jd^3x-\int_\Omega A^{k}u\partial_k\phi_jd^3x \\
&+\oint_{\partial\Omega} A^{k}u \phi_j n_k\frac{1}{\sqrt{\gamma}}d^2\Sigma +\int_\Omega S \phi_j d^3x \label{eq: weak2},
\end{split}
\end{align}
where the mass matrix $M_{ij}$ is defined as
\begin{align}
M_{ij}=\int_\Omega\phi_i\phi_jd^3x.
\end{align}

We solve Eq.~\eqref{eq: weak2} to get $\partial_tu_i$, and use total variational diminishing (TVD) third order Runge-Kutta method \cite{10.2307/2584973} to update $u_i$ in time. Following \cite{Teukolsky:2015ega}, we will work with a nodal expansion (in this work, we use Lagrange interpolating polynomials as our basis and test functions), which is an interpolation for some choice of grid points $x_i$. For simplicity, we choose the grid nodes $x_i$ to be the quadrature nodes and evaluate the integrals with quadrature rules on triangles and tetrahedra from \cite{zhang2009set}. For a nonlinear term like $A^ku$, it is expanded as
\begin{align}
A^ku=\sum_jA^k_ju_j\phi_j
\end{align}
where $A^k_j$ and $u_j$ are the values at the grid nodes of function $A^k$ and $u$. This expression is not exact and will introduce aliasing error, which could make our simulation unstable. Filtering is required to get rid of aliasing error. We can construct Legendre polynomials and Vandermonde matrix for each tetrahedral element using the method introduced by \cite{Hesthaven2008Nodal} and filter the higher modes in the solution’s modal representation. However, the filtered solution will be discontinuous at the boundaries between every two elements. This discontinuity might not be a problem for discontinuous Galerkin (DG) finite element method, but it will invalidate our continuous Galerkin (CG) method. Instead, we use the filter developed by Fischer and Mullen \cite{FISCHER2001265,PASQUETTI2002646},
\begin{align}
F_{\alpha}=\alpha F_{N-1}+(1-\alpha)Id, \label{eq:filter}
\end{align}
where $F_{N-1}$ is the interpolation operator from the space of the polynomials of maximum degree $N$ to the space of the polynomials of maximum degree $N-1$, $Id$ is the identity operator, and $\alpha\in (0,1]$ is the relaxation parameter which allows us to filter only a fraction of the highest mode. Since this filter is based on interpolations in physical space, the filtered solution will still be continuous at the boundaries between every two elements.

\section{\label{sec:level1_3}Numerical Results}

\subsection{Nonlinear scalar waves}
For the code test, we investigate a nonlinear wave equation first. The nonlinear term that we add is the same as the one in \cite{Tichy:2006qn}. The evolution equations in Cartesian coordinate can be written as
\begin{align}
\partial_t\psi=&\Pi \label{eq:wave1}\\
\partial_t\Pi=&\partial_k\Phi_k-\lambda\frac{\psi^3}{1+\psi^2} \label{eq:wave2}\\
\partial_t\Phi_i=&\partial_i\Pi \label{eq:wave3}
\end{align}
where $\psi$ is the scalar field, $\Pi$ and $\Phi$ are its time derivative and spatial derivative, $\lambda$ parametrizes the nonlinearity. The characteristic fields for the system, which are associated with the outward directed unit normal $n_i$ to the boundary, are
\begin{align}
{u}^{{\hat 0}}   &= \psi, \quad & \mathrm{speed}&\,\,  0, \\ 
{u}^{\hat 1\pm}     &= \Pi\mp n_k\Phi_k, & \mathrm{speed}&\,\, \pm 1, \\
{u}_i^{\hat 2}    &= \Phi_i-n_in_k\Phi_k, & \mathrm{speed}& \,\, 0.
\end{align}
It is well known that this system is well-posed if boundary condition is of the following form \cite{Gundlach:2004ri,Tichy:2006qn}
\begin{align}
u^{\hat{1}-}=\kappa u^{\hat{1}+}+f, \label{eq:BCscalar}
\end{align}
where $|\kappa|\le1$ and $f$ is a given function. We set $f=0$ for convenience and consider two different cases: $\kappa= -1$ and $\kappa=0$, which correspond to reflect and radiative boundary conditions, respectively. 

The initial data is chosen to be a Gaussian wave package, 
\begin{align}
\psi=&\frac{\epsilon}{r} \exp\left[-\frac{\tan^2(\frac{\pi(r-R_{in})}{R}-\frac{\pi}{2})}{(\sigma/R)^2}\right], \\
\Pi=&0,
\end{align}
where $R=R_{out}-R_{in}$, $R_{out}$ and $R_{in}$ are the outer and inner boundaries of our computational domain, $\epsilon$ and $\sigma$ are two parameters which characterize the amplitude and width of the wave package correspondingly. In the simulations, we set $\epsilon=0.5, \sigma=4, R_{in}=1.8, R_{out}=11.8$.

\begin{figure}
\includegraphics[scale=0.17,clip=true]{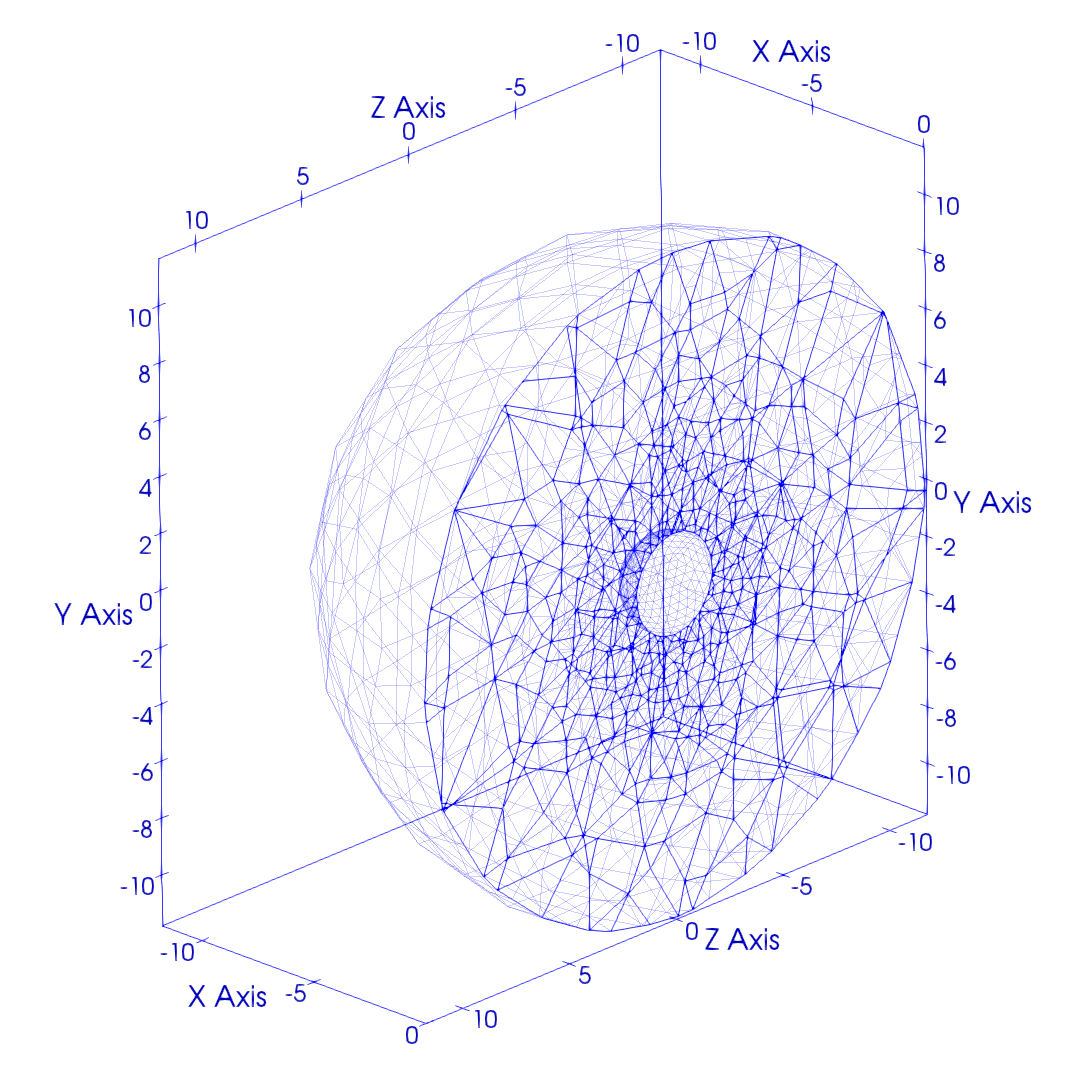} 
\caption{The initial grid structure in the plane $x=0$.}\label{fig:mesh}
\end{figure}

We use NetGen to generate a hollow spherical polyhedron shell grid with radius $r\in[R_{in},R_{out}]$, which is made up of 1235 vertices and 6444 simplest tetrahedral elements, see Fig.~\ref{fig:mesh}. Then we uniformly refine the grid $N$ times. After that, we refine the boundary elements three times. At both the inner and outer boundaries, we set the boundary conditions according to Eq.~\eqref{eq:BCscalar}. For the radiative boundary condition or $f=\kappa=0$ case, we are basically  ``freezing" the ingoing mode $u^{\hat{1}-}$ to its initial value, which is zero.

\begin{figure}
\includegraphics[scale=0.55,clip=true]{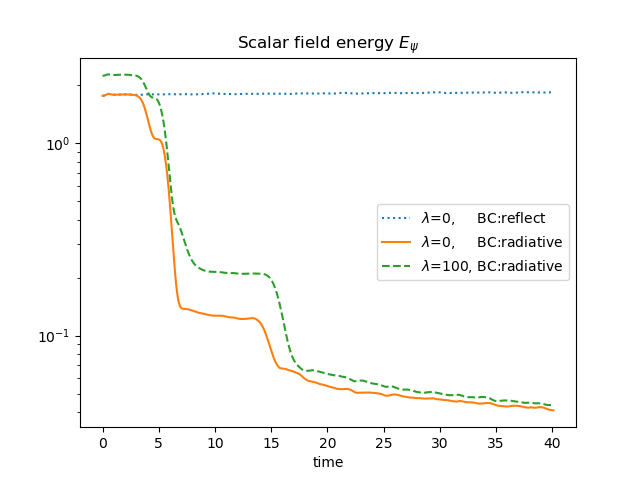} 
\caption{The scalar field energy of three different cases. No energy can escape the computation domain if we implement a reflect boundary condition (dotted line). If we use the radiative boundary condition, the scalar leaves the domain and the energy drops. The linear ($\lambda=0$, broken line) and nonlinear ($\lambda=100$, solid line) cases have the same qualitative behavior. We have used first order polynomials in all three cases.}\label{fig:E_L1}
\end{figure}

Fig.~\ref{fig:E_L1} shows the time evolution of scalar field energy $E_\psi=\int_\Omega \rho d^3x$ with the energy density
\begin{align}
\rho=\frac{\Pi^2}{2}+\frac{\Phi_i^2}{2}+\frac{\lambda}{2}\left(\psi^2-\log(1+\psi^2)\right).
\end{align} 
Three cases with different boundary conditions and nonlinearity are presented. As we can see, the energy is conserved for the $\lambda=0$ case with reflecting boundary condition. While for the $\lambda=0$ case with the radiative boundary condition, the energy drops when the scalar field leaves the grid. The qualitative behavior of the $\lambda=100$ case with radiative boundary condition is the same with the corresponding $\lambda=0$ case, which means that the radiative boundary condition can also propagate nonlinear waves off the grid.

\begin{figure}
\includegraphics[scale=0.5,clip=true]{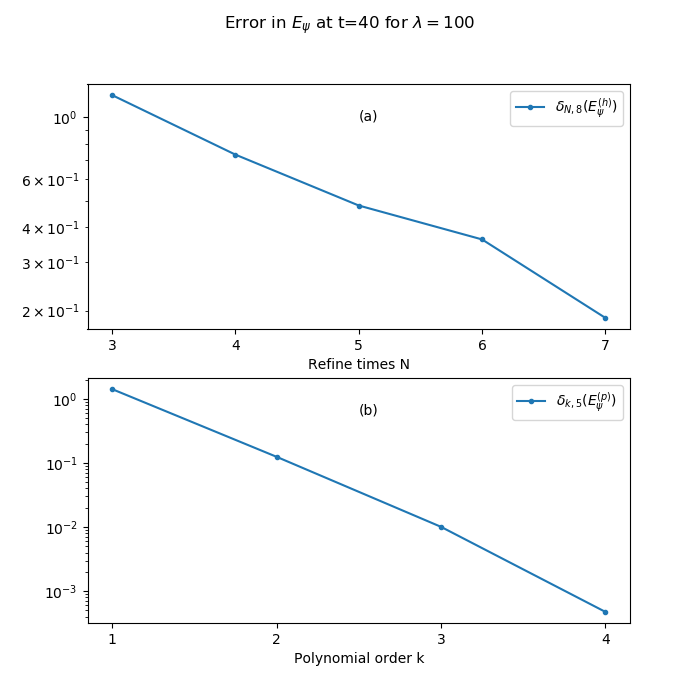} 
\caption{The error in energy at $t=40$ for the nonlinear wave equation with radiative boundary condition. It decreases in both cases as we uniformly refine the grid more times (upper panel) or use higher order polynomials (lower panel).}\label{fig:Cvg_wav}
\end{figure}

There are two kinds of refinements that we can do to improve the accuracy of solutions: $h$-refinement (by further splitting each element) and $p$-refinement (by increasing the order of polynomial basis). We have tried both methods and obtained convergence under both refinements. In Fig.~\ref{fig:Cvg_wav} (a) we show the convergence under $h$-refinement. The error in energy is defined as 
\begin{align}
\delta_{N,8}{(E^{(h)}_\psi)}=|E^{(h)}_{\psi, N}-E^{(h)}_{\psi, 8}|,
\end{align}
where $E^{(h)}_{\psi,N}$ is the scalar field energy measured at $t=40$, with the grid generated from the initial grid by $N$ times uniform refinement. We use $E^{(h)}_{\psi, 8}$ as the reference energy. The basis functions used here are first-order Lagrange polynomials (P1). As we can see, the errors decrease as we split each element.

Fig.~\ref{fig:Cvg_wav}(b) illustrates the convergence under $p$-refinement, as we increase the order of polynomial basis. Similarly, the error in energy is defined as 
\begin{align}
\delta_{k,5}{(E^{(p)}_\psi)}=|E^{(p)}_{\psi, k}-E^{(p)}_{\psi, 5}|,
\end{align}
where $E^{(p)}_{\psi,k}$ represents scalar field energy obtained at $t=40$, using $k$-th order Lagrange polynomials as the basis functions. The reference energy is $E^{(p)}_{\psi,5}$. The grid used in this subplot is obtained from the initial grid after three times uniformly refinement. As expected, the errors decrease exponentially with the order $k$.

\subsection{Schwarzschild black hole in Kerr-Schild coordinate}
Now we turn to evolve the spacetime of a single black hole. As initial data, we use the metric of a Schwarzschild black hole in Kerr-Schild coordinates \cite{Baumgarte:1385040},
\begin{align}
g_{ab}=\eta_{ab}+\frac{2M}{r}l_al_b, \label{eq:KerrSchild}
\end{align}
where $\eta_{ab}$ is the Minkowski metric, $M$ is the mass of the black hole. In Cartesian coordinates, $r=(x^2+y^2+z^2)^\frac{1}{2}$ and $l_a=\left(1,\frac{x_i}{r}\right)$. We use the units where $M=1$.

We numerically evolve Eq.~\eqref{eq: Einsteins} (or Eq.~\eqref{eq: weak2}) using continuous Galerkin finite element method that we described in Sec.~\ref{sec:level1_2}. The gauge source function is initialized based on the metric \eqref{eq:KerrSchild}, which is left constant during the simulation \cite{Bruegmann:2011zj},
\begin{align}
H_a(t=0)=-\Gamma_a(t=0),\quad \partial_tH_a=0.
\end{align}

The characteristic fields for the first-order GH system, Eq.~\eqref{eq: Einsteins}, are given by (c.f., Eq.~(32-34) of \cite{Lindblom:2005qh})
\begin{align}
{u}^{{\hat 0}}_{ab}   &= g_{ab}, \quad & \mathrm{speed}&\,\,  -(1+\gamma_1)n_k\beta^k, \\ 
{u}^{\hat 1\pm}_{ab}     &= \Pi_{ab}\pm n^i\Phi_{iab}-\gamma_2g_{ab}, & \mathrm{speed}&\,\, -n_k\beta^k\pm \alpha, \\
{u}^{\hat 2}_{iab}    &= P_i{}^k\Phi_{kab}, & \mathrm{speed}& \,\, -n_k\beta^k,
\end{align}
which are associated with the outward directed unit normal $n_i$ to the boundary. The projection operator is defined as $P_i{}^k=\delta_i{}^k-n_in^k$. Our computational domain consists of a hollow spherical polyhedron shell that extends from $r_{min}=1.8$ to $r_{max}=11.8$, see Fig.~\ref{fig:mesh}. The inner boundary is lightly inside the horizon which is located at $r_{EH}=2$. At the inner boundary, all the characteristic modes are outgoing (relative to the computational domain), so no boundary condition needs to be imposed. At the outer boundary, we  ``freeze" the values of incoming characteristic fields to their initial values \cite{Lindblom:2005qh,Bruegmann:2011zj}.

Again, we generate the initial mesh with the simplest tetrahedral element decomposition with 1235 vertices and 6444 elements using NetGen. Then we use the dimensionless constraint $L_2$ norm over each element
\begin{align}
||\mathcal{C}^{(i)}||/||\partial \mathcal{U}^{(i)}||=\sqrt{\frac{\int_{\text{i-th element}} \mathcal{C}^2\sqrt{g}d^3x}{\int_{\text{i-th element}} \partial \mathcal{U}^2\sqrt{g}d^3x}},
\end{align}
as the refinement indicator and let PHG do the adaptive refinement, until some preset threshold for the dimensionless constraint $L_2$ norm over the whole computational domain ${||\mathcal{C}_{0}||}/{||\partial\mathcal{U}_{0}||}$ is met. Here $\mathcal{C}$ is a measure of the constraint violations, and $\partial\mathcal{U}$ is a measure of the first order derivatives (c.f., Eq.~(A.2) and Eq.~(A.3) of \cite{Rinne:2007ui}).

\subsubsection{Filtering}
As we have mentioned in Sec. \ref{sec:level1_2}, we used a filter \eqref{eq:filter} to control the aliasing error. To understand this error, let's consider, for example, the integral of two functions, $\int_\Omega f(\bm{x})g(\bm{x}) d^3x$, which are both defined on the domain $\Omega$. Suppose that the quadrature rule we use is $k$-th order, then the polynomials used to expand $f$ and $g$ are also $k$-th order since we have chosen the grid nodes to be the quadrature nodes for convenience. However, when the functions are expanded using $k$-th order basis, the proper quadrature rule should be $2k$-th order. If we still integral the product of these two functions using $k$-th order quadrature rule, it will not be exact. Those modes with order higher than $k$-th will not be well resolved and be `aliased' into lower order modes. To improve this error, we can, of course, prepare another $2k$-th order quadrature rule. But the algorithm will lose the convenience and be very expansive for any realistic, long-term simulations. 

\begin{figure}
	\includegraphics[scale=0.55,clip=true]{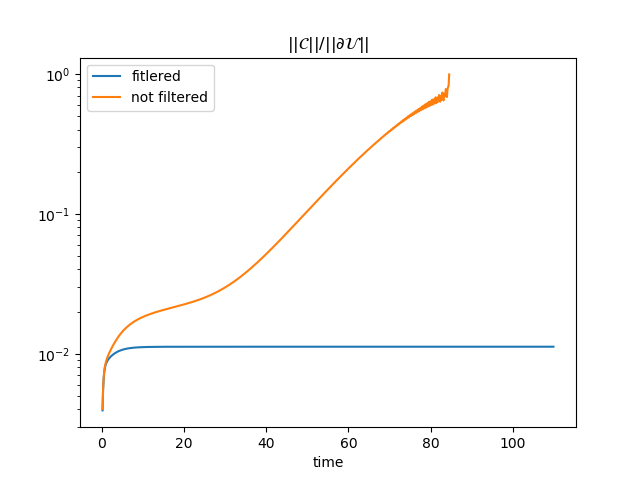}
	\caption{Evolution of dimensionless constraint violations with and without a filter.}\label{fig:filter}
\end{figure}

Instead, we address the aliasing error by filtering a fraction of the highest modes in physical space using Eq.~\eqref{eq:filter}, where $\alpha$ controls the strength of the filter's effect. The filter is applied after each full time step. And it turns out to be a crucial ingredient for numerical stability. In Fig.~\ref{fig:filter}, we plot the dimensionless constraint violations of two simulations with and without filtering. If the solution is not filtered after each time step, the constraint violation brows up after a three-stage evolution: after an initial increase, it settles down for a little while, and finally it starts to grow exponentially without bound at $t\simeq 20$. However, with filtering, the dimensionless constraint violation becomes flat after an initial increase, which indicates that the system becomes stable after filtering a fraction of the highest modes in physical space. Here we used $4$-th order Lagrange polynomials (P4) as basis functions.

\subsubsection{Boundary condition implementation}
The boundary condition is also a vital ingredient for numerical stability. There exists a number of sophisticated and complicated outer boundary condition for the GH system, see \cite{Lindblom:2005qh,Ruiz:2007hg,Rinne:2007ui,Rinne:2008vn}. However, since our focus here is on exploring the finite element as a mean of solving the Einstein equations on unstructured (tetrahedral) grids, we ignore these boundary conditions and use the simplest condition that is successful for the single black hole test case:  ``freezing" the incoming characteristic fields \cite{Lindblom:2005qh,Bruegmann:2011zj},
\begin{align}
\left.\partial_t{u^{\hat{\alpha}}}\right|_{\text{boundary}}=0\quad \text{for}\quad v_{(\hat{\alpha})}<0, \label{eq:BCstrong1}
\end{align}
where $v_{(\hat{\alpha})}$ is the characteristic speed.

We can transform Eq.~\eqref{eq:BCstrong1} back to the condition regarding primitive valuables and impose them on each boundary grid node,
\begin{align}
\partial_tu_i=B_i, \label{eq:BCstrong2}
\end{align}
where $B_i$ represent the boundary condition evaluated at the boundary node $\bm{x}_i$. At the outer boundary, the state vector $u^\alpha$ is integrated in time using Eq.~\eqref{eq:BCstrong2}, instead of using Eq.~\eqref{eq: weak2}. Unfortunately, the runs which impose the outer boundary condition in this way are not stable. The reason we suspect for the instability is the inconsistency between the weak form evolution in the bulk \eqref{eq: weak2} and the strong form evolution at the outer boundary \eqref{eq:BCstrong2}. 

Therefore, we modify the form of boundary condition \eqref{eq:BCstrong1} by integrating it in time and obtain
\begin{align}
\left.u^{\hat{\alpha}}\right|_{\text{boundary}}=C^{\hat{\alpha}}\quad \text{for} \quad v_{(\hat{\alpha})}<0, \label{eq:BCweak1}
\end{align}
where $C^{\hat{\alpha}}$ is constant in time and determined by initial data. We again transform Eq.~\eqref{eq:BCweak1} back to the condition concerning primitive valuables
\begin{align}
u_i=B_i^{(C)}, \label{eq:BCweak2}
\end{align}
where $B_i^{(C)}$ is the boundary condition regarding state vector. Then the ``freezing"  incoming characteristic fields boundary condition can be imposed in weak form as follows: at each time step, we replace the primitive valuables $u^\alpha$ present in the surface integral terms in Eq.~\eqref{eq: weak2} with $B_i^{(C)}$. This weak form boundary condition works well and removes the instability present in the strong form boundary condition cases.

\begin{figure}
	\includegraphics[scale=0.55,clip=true]{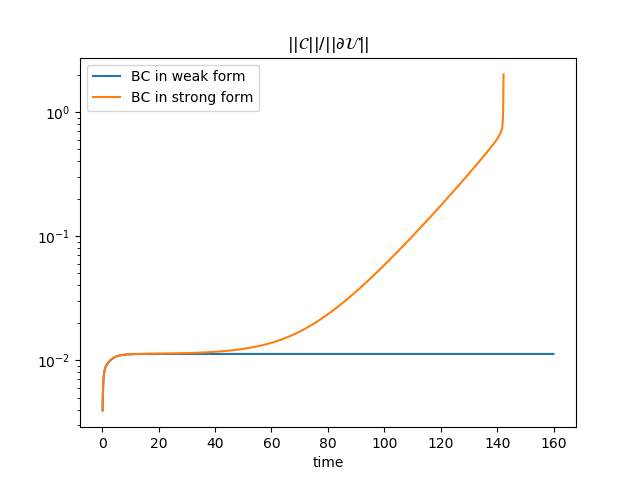}
	\caption{Evolution of dimensionless constraint violations with outer boundary conditions implemented using strong and weak forms}\label{fig:strong}
\end{figure}

In Fig.~\ref{fig:strong} we show two simulations with their outer boundary conditions imposed using strong and weak forms. They share the same behavior before $t\simeq 20$: after an initial increase, the dimensionless constraint violations become flat for a while. Then the strong form case diverges from the weak form case and increases exponentially until the run fails. 

\begin{figure}
	\includegraphics[scale=0.55,clip=true]{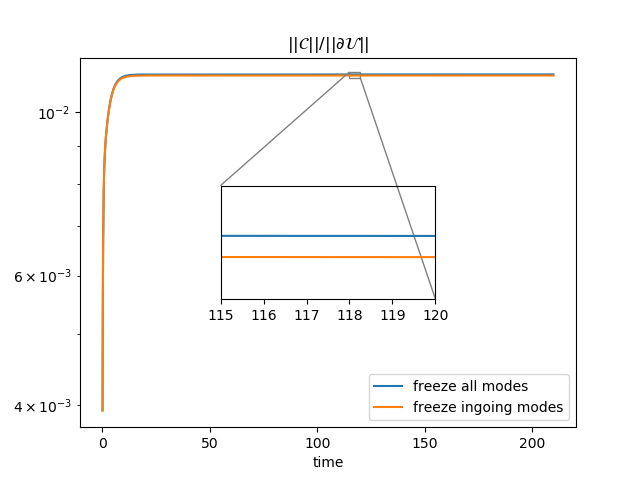}
	\caption{Evolution of dimensionless constraint violations with  ``freezing"  all modes and  ``freezing"  only ingoing modes at outer boundary.}\label{fig:freeze}
\end{figure}

Our code is still stable if we  ``freeze" all the modes on the outer boundary, or in other words, fix the boundary value to the analytic solution, just as we found in \cite{Cao:2018myc}. We plot in Fig.~\ref{fig:freeze} the result of two cases where we  ``freeze" all the modes and  ``freeze" only the ingoing modes. As we can see, the behaviors of the dimensionless constraint violations are almost the same for these two cases, except that the   ``freezing"  all modes case settles down to a little bit larger value of constraint violation.

\subsubsection{Convergence}

\begin{figure}
\includegraphics[scale=0.5,clip=true]{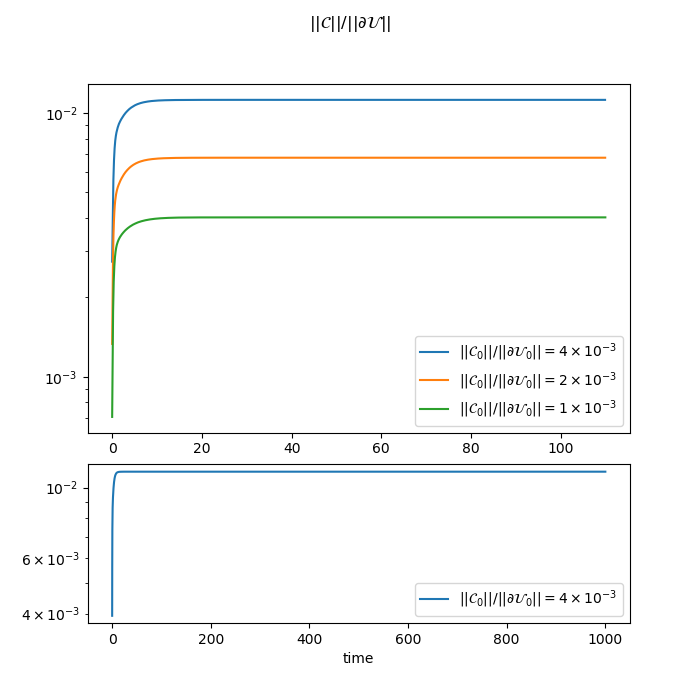}
\caption{Evolution of dimensionless constraint violations for Schwarzschild initial data. In the upper panel, we show the evolutions using three different numerical resolutions. The lower panel shows the long timescale evolution for $||\mathcal{C}_0||/||\partial \mathcal{U}_0||=4.0\times 10^{-3}$. }\label{fig:C_L2}
\end{figure}

In Fig.~\ref{fig:C_L2} we show the convergence and stability of the Schwarzschild black hole evolution. The simulations are carried out using 4-th order Lagrange polynomials (P4) as basis functions. The top panel displays the dimensionless constraint violations over the whole computational domain, $||\mathcal{C}||/||\partial \mathcal{U}||$, with different resolutions. The grids used in these three cases are generated with different preset threshold $||\mathcal{C}_0||/||\partial \mathcal{U}_0||$, which labels the resolutions. As we can see, they share the same qualitative behavior: after an initial increase, $||\mathcal{C}||/||\partial \mathcal{U}||$ settles down to a constant. In particular, $||\mathcal{C}||/||\partial \mathcal{U}||$ settles down to $4.0\times 10^{-3}$, $6.8\times 10^{-3}$ and $1.1\times 10^{-2}$, corresponding to the preset threshold $||\mathcal{C}_0||/||\partial \mathcal{U}_0||$ equals $1.0\times 10^{-3}$, $2.0\times 10^{-3}$ and $4.0\times 10^{-3}$. The dimensionless constraint violations decrease as we increase the resolution.

The last case where $||\mathcal{C}_0||/||\partial \mathcal{U}_0||=4.0\times 10^{-3}$ has been evolved to $t=1000$ in the bottom panel of Fig.~\ref{fig:C_L2} and we see no sign of instability. We conclude that our code is convergent and stable up to at least $t=1000$, and, we presume, forever. 

\subsubsection{Strong-scaling}
\begin{figure*}
	\centering
	\includegraphics[scale=0.68,clip=true]{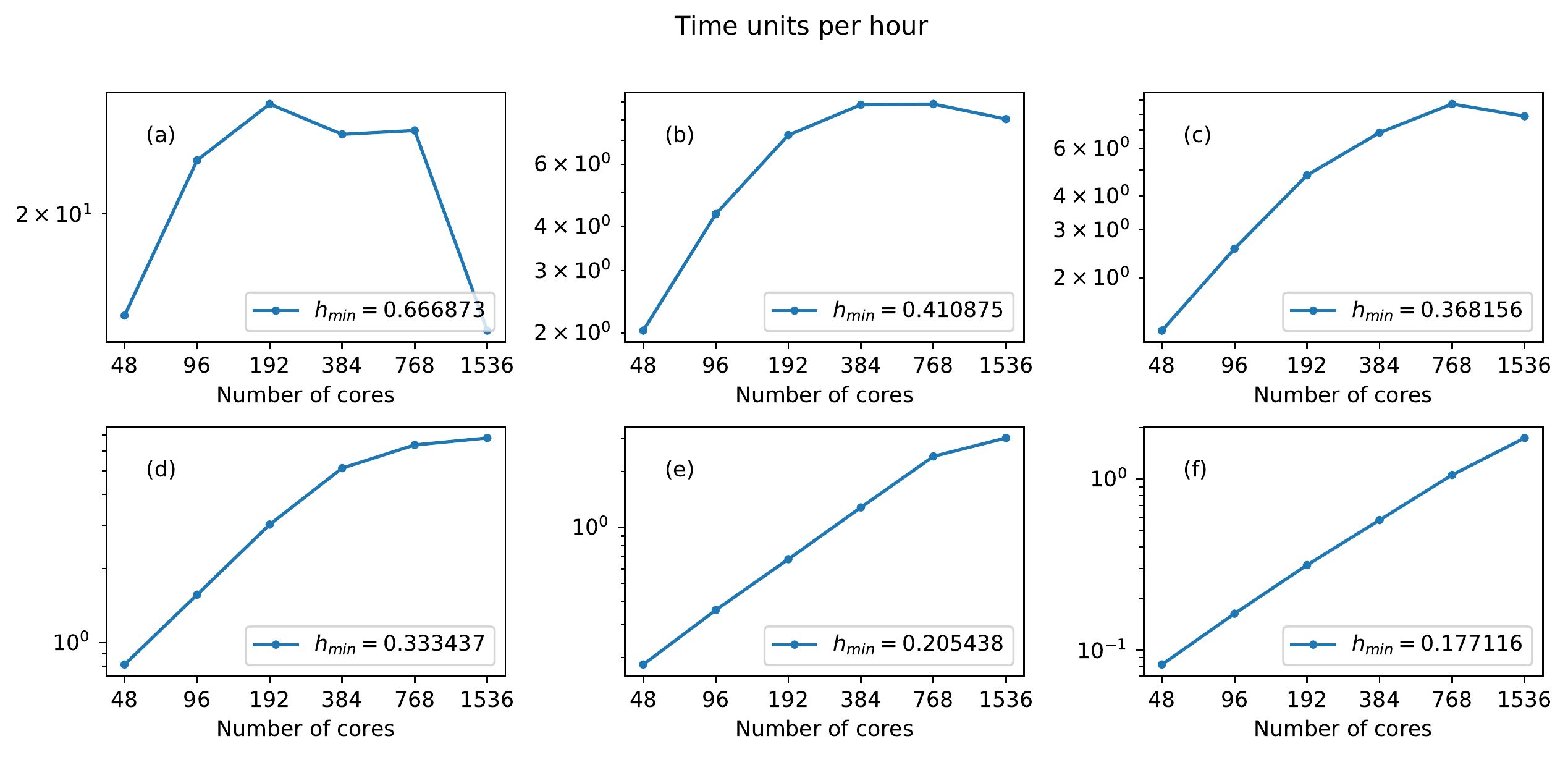}
	\caption{Strong scaling of iPHG with different number of elements on Tianhe-2 cluster. Subplots (a)-(f) correspond to tests whose meshes are refined $0-5$ times from the initial mesh. We use the diameter  of the smallest element $h_{min}$ to label the resolutions of different meshes.}\label{fig:strong_scaling}
\end{figure*}

In Fig.~\ref{fig:strong_scaling} we display the strong scaling plots performed on the Tianhe-2 (NSCC-GZ) cluster located at Sun Yat-sen University, with Intel Xeon E5-2692 v2 processors. Our strong scaling tests are performed by evolving single black hole spacetime on meshes with different resolutions, which are labelled by the diameters of the smallest elements. From subplot (a) to (f), the meshes are generated from the initial mesh (with simplest tetrahedral elements decomposition with 1235 vertices and 6444 elements) through zero to five times uniform refinement. In subplot (a), we observe that the strong scaling breaks down when more than 192 cores are used, which means iPHG can only use as much as 192 cores effectively in this case. However, as the resolutions (or the number of elements) are increased, the inflection point of strong scaling moves gradually right to larger number of cores. In subplot(f), the inflection point disappears and iPHG can effectively use more than 1536 cores, which reflects perfect scaling. 

For an IMRI, which is a significant target of iPHG, there exists a massive difference between the small-scale dominated by the size of the smaller black hole and the large-scale dominated by the range of the whole binary system. We need not only to resolve the small-scale, but to cover the entire range of the system, which makes this a very challenging computational problem. Even though we may alleviate it through highly effective adaptive mesh refinement, a large-sized calculation is still inevitable. Therefore, we need our code to have good parallel scalability. 

For the small-sized calculation, the parallel scalability of iPHG is not very good, see subplot (a) of Fig.~\ref{fig:strong_scaling}. But its performance become better and better with the increase in calculation size (or the number of elements), as we can see in Fig.~\ref{fig:strong_scaling}. We can expect iPHG to have good parallel scalability when it is used to simulate IMRIs since the calculation size of IMRI is inevitably huge.

\section{\label{sec:level1_4}Summary and discussion}
\label{discussion}
A new finite element code, Einstein PHG (iPHG), has been extended to solve the evolution part of Einstein equations in first-order GH formalism. It has two main features: first, thanks to PHG, iPHG have good parallel scalability, which is very crucial for the simulations of IMRIs. Second, it is equipped with unstructured mesh and can do parallel adaptive mesh refinement and coarsening, which we believe will benefit a lot when it is used to simulation the binary neutron star coalescence or black hole-neutron star coalescence.

As a first step, we applied iPHG to evolve the spacetime of a single black hole. Before going to the Einstein equations, we tested our code by solving nonlinear wave equations. Our code worked well with both reflect and radiative boundary conditions and exhibited both convergences with h-refinement and p-refinement. For the single black hole case, we found that filter and boundary conditions were both crucial ingredients for numerical stability. We armed iPHG with the filter \eqref{eq:filter} developed by Fischer and Mullen. For simplicity, the  ``freezing"  incoming characteristic fields condition was imposed in weak form at the outer boundary. We showed that the algorithm is convergent and stable for long-timescale spacetime evolution.

In future work, we intend to combine our algorithm with some discontinuity capturing schemes \cite{Hughes1986A,Tezduyar1986Discontinuity,John2007On,John2008On,Codina1993A} to suppress oscillations that may occur near shocks, such that it can handle coupled Einstein equations with hydrodynamics equations. We would also like to explore the discontinuous Galerkin method as a mean of solving Einstein and hydrodynamics equations, just as in \cite{Dumbser:2017okk,Hebert:2018xbk}, but using the unstructured grid.

\begin{acknowledgments}

We thank David Hilditch, Scott Field, Lee Lindblom,  Wolfgang Tichy, Yun-Kau Lau for helpful discussions.  We are very grateful to Lin-Bo Zhang for helping with PHG usage. This work was supported by the National Natural Science Foundation of China Grants No.11690022, No. 11435006, No.11447601 and No.11647601, and by the Strategic Priority Research Program of CAS Grant No.XDB23030100, and by the Peng Huanwu Innovation Research Center for Theoretical Physics Grant No.11747601, and by the Key Research Program of Frontier Sciences of CAS. 

\end{acknowledgments}




\bibliography{refs}

\end{document}